\documentclass[twocolumn,amsmath,amssymb,superscriptaddress]{revtex4-1}

\usepackage{color}% color text
\usepackage{graphicx}
\usepackage{amssymb}
\usepackage{amsmath}
\usepackage{mathrsfs}
\newcommand{\ud}{\mathrm{d}}

\newcommand{\pdern}[3]{\frac{\partial^#3#1}{\partial#2^#3}}
\newcommand{\pder}[2]{\frac{\partial#1}{\partial#2}}
\let\Eps\varepsilon
\usepackage[breaklinks=true]{hyperref}
\graphicspath{ {./figures/} {./}}

\begin{document}

\title{Nonlinear stage of Benjamin-Feir instability in forced/damped deep-water waves}

\author{Andrea Armaroli}
\email{andrea.armaroli@unige.ch}
\affiliation{GAP, Universit{\'e} de Gen{\`e}ve, Chemin de Pinchat 22, 1227 Carouge, Switzerland}
\affiliation{Institute for Environmental Sciences, Universit{\'e} de Gen{\`e}ve, Boulevard Carl-Vogt 66, 1205 Gen{\`e}ve, Switzerland}
\author{Debbie Eeltink}
\affiliation{GAP, Universit{\'e} de Gen{\`e}ve, Chemin de Pinchat 22, 1227 Carouge, Switzerland}
\affiliation{Institute for Environmental Sciences, Universit{\'e} de Gen{\`e}ve, Boulevard Carl-Vogt 66, 1205 Gen{\`e}ve, Switzerland}

\author{Maura Brunetti}
\affiliation{GAP, Universit{\'e} de Gen{\`e}ve, Chemin de Pinchat 22, 1227 Carouge, Switzerland}
\affiliation{Institute for Environmental Sciences, Universit{\'e} de Gen{\`e}ve, Boulevard Carl-Vogt 66, 1205 Gen{\`e}ve, Switzerland}

\author{J{\'e}r{\^o}me Kasparian}
\affiliation{GAP, Universit{\'e} de Gen{\`e}ve, Chemin de Pinchat 22, 1227 Carouge, Switzerland}
\affiliation{Institute for Environmental Sciences, Universit{\'e} de Gen{\`e}ve, Boulevard Carl-Vogt 66, 1205 Gen{\`e}ve, Switzerland}
\date{\today}

\begin{abstract}
We study a three-wave truncation of a recently proposed damped/forced high-order nonlinear Schr{\"o}dinger equation (DF-HONLS) for deep-water gravity waves under the effect of wind and viscosity. The evolution of the norm (wave-action) and spectral mean of the full model are well captured by the reduced dynamics. Three regimes are found for the wind-viscosity balance: we classify them according to the attractor in the phase-plane of the truncated system and to the shift of the spectral mean. A downshift can coexist with both net forcing or damping, i.e. attraction to period-1 or period-2 solutions. Upshift is associated to stronger winds, i.e.~to a net forcing where the attractor is always a period-1 solution. The applicability of our classification to experiments in long wave-tanks is verified.
\end{abstract}

\maketitle

\section{Introduction}

The development of  accurate  propagation equations for water waves in the ocean has benefitted from the evolution of computational capacity and the abundance of experimental data from human activities in the oceans (e.g.~oil platforms) and satellites.  The study of the role of different physical effects, particularly in relation to the debate about the origin and explanation of rogue waves\cite{Akhmediev2009,Onorato2013,Birkholz2015}, has a practical impact on the forecast and prevention of catastrophic  ship accidents and a fundamental interest in  understanding  the limits of applicability of the current approximated models.

Naturally, the first step is to include in one dimensional propagation equations the two main forcing and damping mechanisms, viscosity and wind.
%, and to compare the improved predictions to experiments in water tanks. 
The former is an unavoidable limit in laboratory facilities, the impact of which on propagation is non-trivial \cite{Segur2005,Wu2006,Dias2008,Schober2015,Carter2016, Kimmoun2016}; the latter \cite{Miles1957,Hara1991} has attracted much interest in theoretical and experimental efforts, particularly in recent years \cite{Touboul2010,Onorato2012,Brunetti2014,Brunetti2014b,Slunyaev2015a,Toffoli2017}. 

The  nonlinear Schr{\"o}dinger equation (NLS), a universal model used, among others, to describe the propagation of gravity waves in deep-water, can be extended so as to include forcing and damping terms which describe the effect of wind and viscosity in one-dimensional propagation \cite{Eeltink2017}.
Forcing and damping can be decomposed in a homogeneous contribution at the same order (in the wave steepness) of the conventional NLS and a frequency-dependent (dispersive) higher-order term, which is thus in direct competition with the Dysthe higher-order corrections to NLS (HONLS) \cite{Dysthe1979}. 
Some experimental results in wave tanks, where spectral asymmetry develops, can be thus better understood. Consider, for example, the Benjamin-Feir instability (BFI), the growth of modulated waves on top of a uniform carrier wave \cite{Benjamin1967, Zakharov2009, Henderson2012a}: earlier experiments showed indeed a downshift of the frequency peak of the wave \cite{Lake1977}. 
%Much modeling effort was devoted to explain it.

It was theoretically demonstrated that viscosity stabilizes BFI \cite{Segur2005,Wu2006} and can lead to a  {spectral} downshift \cite{Carter2016}, while wind can destabilize even modes outside the conventional BFI band in an asymmetric fashion \cite{Brunetti2014}. The experimental assessment is complicated by the limited length of available wave-tanks, inaccurate estimates of viscosity, turbulent wind-water interaction and  wave-breaking of large waves.  It seems clear, however, that the interplay of wind  and viscosity can lead to both up- and down-shift \cite{Eeltink2017}.

We remark here that the definition of spectral shift can refer to peak frequency \cite{Zavadsky2013} or to spectral mean \cite{Carter2016,Eeltink2017}, the former is experimentally more practical, but the latter is certainly more scientifically sound \cite{Armaroli2017}. In fact the HONLS for the envelope of the free-surface elevation  conserves only the norm and  predicts a spectral peak downshift, albeit not permanent; the spectral mean of its solutions exhibits instead  a temporary upshift. This is a consequence of the breaking of the translational symmetry of the system (i.e.~the momentum is not conserved), which can be restored by a different choice of variables \cite{Lo1985,Gramstad2011,Fedele2012}. These canonical variables involve not only the envelope of the surface elevation but also the velocity potential and thus do not correspond to experimentally accessible quantities, for which the HONLS proves effective  \cite{Shemer2002,Shemer2010,Chabchoub2013}.

%It seems clear, however, that the correct definition of spectral shift must rely on the spectral mean and in this framework, the wind interplay with viscosity can lead to both up- and down-shift \cite{Eeltink2017}. It is also clear that forcing and damping lead to a change in the topology of solutions with respect to the nearly-integrable HONLS system. 

Finally, we should not forget that exact solutions, such as the Akhmediev breather (AB) \cite{Akhmediev1986,Akhmediev1987b}, are not available in the HONLS case; we   rely instead on a three-mode truncation  \cite{WhithamBook,Trillo1991c,Armaroli2017}. This approach provides good approximated solutions and allows a more flexible choice of initial conditions and range of parameters  than conventional approximated ABs \cite{Zakharov2010,Onorato2012,Brunetti2014b}.

We recently showed \cite{Armaroli2017} that the phase-plane topology in conjunction with the unbalance of BFI sidebands provide, for a single unstable mode, a clear and simple classification of the solutions of HONLS and the associated temporary spectral upshift. 

Here we apply the same approach to the forced/damped equation presented in Ref.~\cite{Eeltink2017} and explicitly derive a system of ordinary differential equations (ODEs) correctly describing the nonlinear behavior of the BFI in that model.
% for the sideband amplitude, the carrier-sideband relative phase, total energy and sideband unbalance. 
We demonstrate that the topology of orbits in the reduced phase space is a complementary criterion (besides upshift and downshift) to understand the long-distance evolution of water waves under the action of wind and viscosity. 

After presenting the model equation and the allowed regimes (Section II), we recall in Section III the three-wave truncation and its predictions for the conservative case. Next we classify the solutions in terms of phase-plane trajectories and spectral shift in the forced/damped regimes  and discuss the limits and perspectives of our model and analysis (Section IV). Section V is devoted to conclusions.
 
\section{Model equations}
 Consider the following damped/forced HONLS (DF-HONLS)\cite{Eeltink2017} for the space evolution of a narrow-band wave-packet propagating in the $\tilde{x}$ direction under the action of wind and viscosity. It reads as
\begin{multline}
\underbrace{ \frac{\partial A}{\partial \tilde{x}} + i\frac{k_0}{\omega_0^2}  \frac{\partial^2 A}{\partial \tilde{t}^2} + i{k_0^3} A|A|^2}_\mathrm{NLS}\\ =  \underbrace{\frac{k_0^3}{\omega_0}\left(6 |A|^2\frac{\partial A}{\partial \tilde{t}} + 2 A\frac{\partial |A|^2}{\partial \tilde{t}} + 2iA \mathcal{H}\left[\frac{\partial |A|^2}{\partial \tilde{t}}\right]\right)}_\mathrm{{HO \;corrections}} \\ 
\underbrace{-4 \frac{k_0^3}{\omega_0}\nu A - 20 i \frac{k_0^3}{\omega_0^2}\nu\pder{A}{\tilde{t}}}_\mathrm{Viscous\; damping}  
 \underbrace{+ \frac{k_0}{\omega_0}\Gamma_m A  +  4i \frac{k_0}{\omega_0^2}\Gamma_m\pder{A}{\tilde{t}}}_\mathrm{Wind\; forcing} 
 \label{eq:FDHONLSdim}
\end{multline}
where $A(\tilde x,\tilde t)$ is the complex amplitude of the surface elevation [m], $k_0$ is the carrier wavenumber [m\textsuperscript{-1}], $\omega_0=\sqrt{gk_0}$ ($g$ is the standard acceleration due to gravity) is the associated angular frequency in the deep water limit, $\tilde{x}$ and $\tilde{t}$ denote space and time in a frame co-moving at the group velocity $C_g=\omega_0/2k_0$ of the wave.  { $\mathcal{H}[\cdot]$ stands for the Hilbert transform.} $\nu \; \mathrm{[m^{2}/s]}$ is defined as the wave dissipation effects, including the  kinematic viscosity of the fluid, surface impurities and side-wall mechanical friction, while the wind blowing in the direction of the wave propagation leads to a  growth rate  $\Gamma_m\; \mathrm{[s^{-1}]}$ of the wave energy, as predicted by the Miles mechanism \cite{Miles1957}. 
The left-hand side is the conventional NLS, which can be derived from the  Euler equations in the incompressible irrotation limit, by means of the method of multiple scales (MMS) at third order in the steepness $\Eps \equiv k_0 A_0/\sqrt{2}$ ($A_0$ is the reference amplitude of an ideally stable Stokes' wave). The right hand side is composed by  HO corrections (fourth-order)  \cite{Dysthe1979,Janssen1983,Lo1985}, viscous damping---in the limit of small viscosity \cite{Dias2008}, and forcing due to wind---a correction in the pressure balance at the water/air interface \cite{Miles1957,Brunetti2014}. We remark that damping and forcing consist each of a homogeneous (third-order) and a derivative part (fourth-order), see App.~\ref{app:DFdisp}. 

In order to minimize the number of free parameters, we define the adimensional variables  $\tau \equiv \tilde t/T_0$, $\xi\equiv \tilde x/L_0$, $a(\xi,\tau)\equiv A/A_0$ with 
$T_0 = 1/(\omega_0\Eps)$ and $L_0 = 1/(2\Eps^{2}k_0)$
and rewrite \eqref{eq:FDHONLSdim} as
\begin{multline}
\underbrace{\pder{a}{\xi} + i \frac{1}{2} \pdern{a}{\tau}{2} + i|a|^2a }_{\mathrm{NLS}}= 
\underbrace{\delta_0 a  + i\delta_1 \pder{a}{\tau}}_{\mathrm{DF}} \\
+{\Eps} \underbrace{\left\{8|a|^2\pder{a}{\tau} + 2 a^2
	\pder{a^*}{\tau} + 2i	a \mathcal{H}\left[\pder{|a|^2}{\tau}\right]\right\}}_\mathrm{{HO\; corrections}}	
	\label{eq:FDHONLSadim}
\end{multline}
where $\delta_0\equiv\frac{T_0}{2\Eps}(\Gamma_m-4 k_0^2\nu)$ and $\delta_1\equiv2{T_0}(\Gamma_m-5k_0^2\nu)$ are  the coefficient of the 0\textsuperscript{th} and 1\textsuperscript{st}-order damping/forcing contributions (3\textsuperscript{rd} and 4\textsuperscript{th}-order, when referred to the MMS expansion of Euler equations). In the angular frequency domain ($ \pder{}{\tau}\to -i\Omega$),  {we write the damping/forcing terms as
\begin{equation}
g(\Omega)\equiv\delta_0 + \delta_1\Omega.
\label{eq:DFdisp}
\end{equation} }
In the absence of wind ($\Gamma_m=0$), the spectrally homogeneous damping dominates; vice-versa for vanishing viscosity ($\nu=0$), the homogeneous wind forcing prevails. 
Notice, however, that $\delta_0$ and $\delta_1$ are each the balance of wind- and viscosity-related contributions with different coefficients. Thus, in principle, the wind strength can be tuned to obtain  $\delta_0$ and $\delta_1$ of nearly the same magnitude or even $\delta_1$  dominant over $\delta_0$. Similar conditions of wind and viscosity compensating each other have been studied\cite{Touboul2010} and allow one to explore the dispersive nature of damping and forcing. 

Since $\nu\ge 0$ we have that $4\Eps\delta_0\ge\delta_1$, thus we obtain three regimes: (i) $\delta_0<0,\delta_1<0$ with $4\Eps|\delta_0|\le|\delta_1|$  (ii) $\delta_0>0,\delta_1<0$ and (iii) $4\Eps\delta_0\ge\delta_1>0$. We will study in section IV what they lead to.

 {We remark that the approximation \eqref{eq:DFdisp} is obtained from a perturbation series expansion and provides an idealized form of dispersive forcing/damping which is not physical, because forcing and damping diverge  at $\Omega\to\pm\infty$. A more physical parametrization  includes at least a parabolic correction $-\delta_2\Omega^2$ with $0<\delta_2\sim\mathcal{O}(\Eps^2)$, in order to account for viscosity to limit energy growth outside the considered bandwidth. In Refs.\cite{Fabrikant1980, Fabrikant1984} it is shown that a phase transformation permits to put the equation in time-symmetric form: nonlinear damping must be included and the coefficients of non-linear phase modulation and damping combine to give the coefficients of the new model. This transformation is very insightful to get a damped-forced NLS and its solitary waves. In our case we prefer working with the simplest possible nonlinear coefficients (neglecting non-linear contributions of wind and viscosity, certainly of higher-order) and to disregard the parabolic term, which would be of fifth-order in steepness, see above. We verified that, if up to a 5$\%$ extra overall damping is included, its impact is negligible on the results presented below. An estimate of $\delta_2$ is presented in App.~\ref{app:DFdisp}.
}

\section{Three-wave truncation}

We apply to \eqref{eq:FDHONLSadim} the three-wave truncation approach \cite[p. 527]{WhithamBook}  and briefly recall, as a reference, its results \cite{Armaroli2017} in the  case of no damping/forcing.
At the beginning of the wave-tank we excite a homogeneous carrier wave perturbed by a pure oscillation at normalized angular frequency $\Omega$. 
As the normalized $a\sim \mathcal{O}(1)$, we can assume that the initial norm $N\equiv\int_{-\infty}^{\infty}{|a|^2\ud \tau}=N_0=1$, so that the peak BFI gain occurs for $\Omega_\mathrm{M}\approx\pm\sqrt{2N_0}(1-\frac{3}{2}\sqrt{2N_0}\Eps)$ \cite{Dysthe1979,Lo1985},  {closer to the BFI cut-off frequency ($\Omega_c=\sqrt{2N_0}(1-2\sqrt{2N_0}\Eps)$) than to zero.}
Thus, letting $\Omega=\Omega_\mathrm{M}$ only a single mode is unstable and the four-wave mixing products of $\pm\Omega$ are stable:  this represents an approximation of the AB solution for the HONLS. 

We can thus restrict our analysis to a three-mode Ansatz
 \begin{equation}
a(\xi,\tau) = a_0(\xi) + a_1(\xi)e^{-i \Omega \tau} + a_{-1}(x)e^{i \Omega \tau}
\label{eq:3wAnsatz}
\end{equation}
which, inserted in \eqref{eq:FDHONLSadim}, allows us to obtain the following system of complex ODEs:
\begin{widetext}
\begin{equation}
\begin{aligned}
	\dot{a}_0 &= -i\left(\left|a_0\right|^2+2\left|a_1\right|^2+2\left|a_{-1}\right|^2\right)a_0 - 2ia_0^*a_1a_{-1} +\underbrace{\delta_0 a_0}_\mathrm{DF} \\
	&+ \underbrace{i{\Eps\Omega}\left\{-8(|a_1|^2-|a_{-1}|^2)a_0 + 4(|a_1|^2-|a_{-1}|^2)a_0 
	+2\, s \left[(|a_{-1}|^2 + |a_1|^2)a_0 + 2a_0^*a_{-1}a_1\right]\right\}}_\mathrm{HO \; corrections}
\\
		\dot{a}_1 &= i\frac{\Omega^2}{2}a_1 - 
		i\left(\left|a_1\right|^2+2\left|a_0\right|^2
		+2\left|a_{-1}\right|^2\right)a_1-ia_0^2a_{-1}^*+\underbrace{\delta_0 a_1  +  \Omega\delta_1 a_1}_\mathrm{DF}
		\\
		&+\underbrace{ i{\Eps\Omega}
		\left\{-8(|a_0|^2+|a_1|^2)a_1 + 2(|a_1|^2-2|a_{-1}|^2)a_1
		- 2 a_0^2 a_{-1}^*+
		2\,  s\left[\left(|a_0|^2 + 2 |a_{-1}|^2\right) a_1 + a_0^2a_{-1}^*\right] 
	\right\}	}_\mathrm{HO\;corrections}
	\\			
		\dot{a}_{-1} &= i\frac{\Omega^2}{2}a_{-1} - 
		i\left(\left|a_{-1}\right|^2+2\left|a_0\right|^2
		+2\left|a_1\right|^2\right)a_{-1}- ia_0^2a_1^*+\underbrace{\delta_0 a_{-1} - \Omega\delta_1 a_{-1}}_\mathrm{DF}
		 \\
		&+ \underbrace{i{\Eps\Omega}
		\left\{8(|a_0|^2+|a_{-1}|^2)a_{-1} - 2(|a_{-1}|^2-2|a_1|^2)a_{-1} + 2  a_0^2a_1^*
		+ 2\, s \left[\left(|a_0|^2 + 2 |a_1|^2\right) a_{-1} + a_0^2a_1^*\right] 
	\right\}}_\mathrm{HO \; corrections}
\end{aligned}
\label{eq:CMT}
\end{equation}
\end{widetext}
where the dot denotes the derivative with respect to $\xi$ and $s=\mathrm{sign}\,{\Omega}$. The main difference with respect to Eq.~4 of Ref.~\cite{Armaroli2017} is that damping-forcing (DF) appears as a uniform contribution to the carrier wave $a_0$  and an asymmetric contribution to $a_{\pm 1}$,  consistently with the  {DF function in Eq.~\eqref{eq:DFdisp}.}
%%%
% Fig. 1 comparison of phase space attractors
\begin{figure*}[t]
\centering
\includegraphics[width=.80\textwidth]{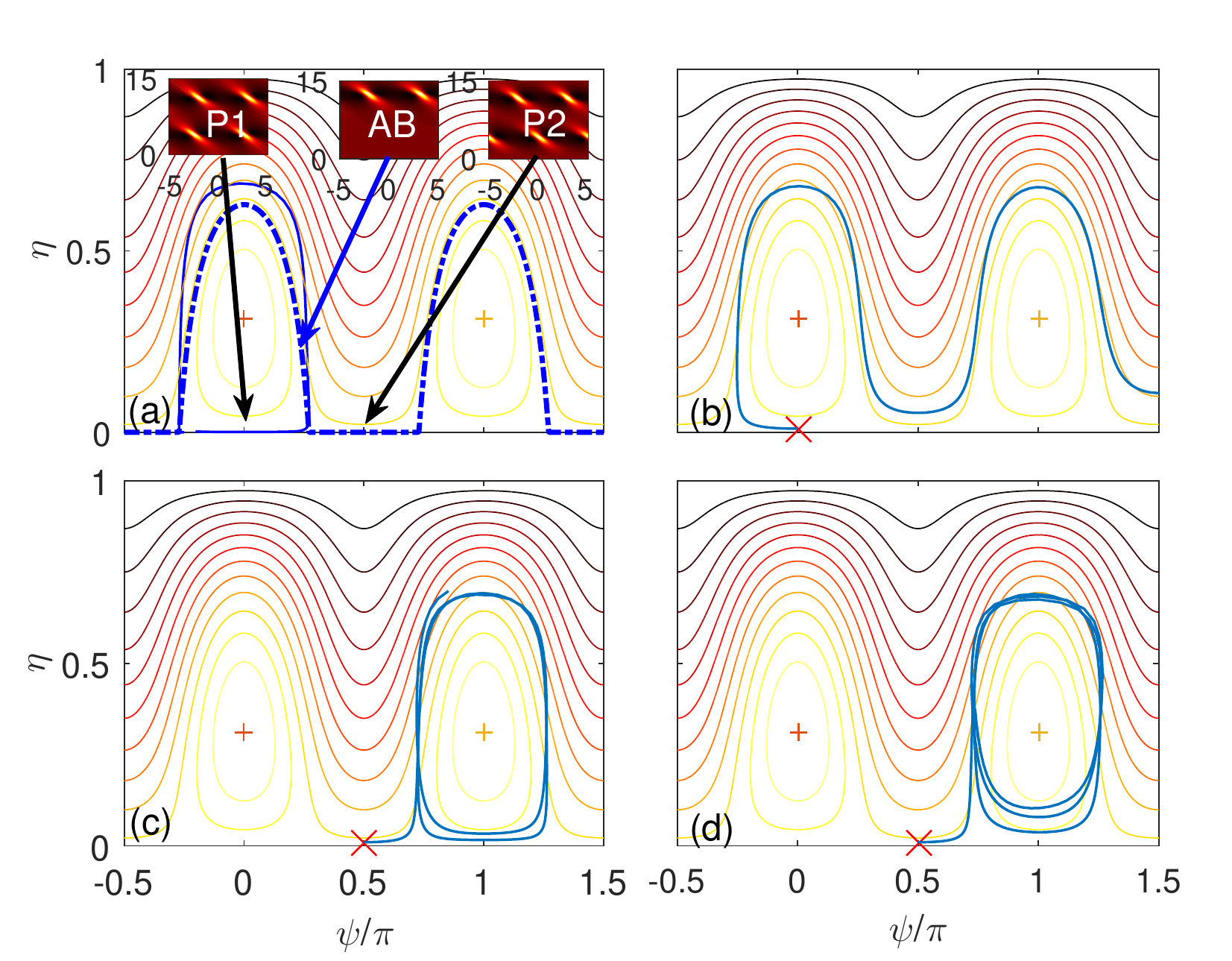}
\caption{Classification of the DF-HONLS solutions in $(\psi/\pi,\eta)$ plane. (a) Conservative case. Main panel:  level sets of the Hamiltonian associated to the integrable version of Eq.~\eqref{eq:CMTreal} (see Appendix \ref{app:Hamiltonian}), for $\Eps=0.1/\sqrt{2}$, $N_3=1$, $\alpha(0)=0$. In the insets the solutions of the HONLS corresponding to three different trajectories in the phase plane: period-one (P1), period-two (P2), Akhmediev breather AB. The latter corresponds to the separatrix in the phase plane (dashed-dotted blue line level set to be compared to actual orbit in solid blue) and is obtained by using as initial condition the AB solution of NLS with $a=0.32$ (corresponding to the peak BFI gain of the HONLS) at $\xi=-10$. Notice that the peak is reached in the HONLS later than in the NLS, as is well known from the literature.  {Trajectories of the HONLS and truncated system qualitatively agree: the topology is equivalent, see text.} (b-d) Representation of the evolution under the effect of wind and viscosity on the phase plane $(\psi,\eta)$. (b)  $\delta_0=-0.01$, $\delta_1=-0.005$, (c) $\delta_0=0.005$, $\delta_1=-0.0025$ (d) $\delta_0=0.01$, $\delta_1=0.0025$. The initial conditions (marked as red crosses) are $\eta(0)=0.01$ for every panel and (b) $\psi(0)=0$, (c-d) $\psi(0)=\pi/2$. We observe that in each case (b-d) the solution is attracted to a family which is different from the one pertaining to its initial conditions.}
\label{fig:phaseplaneGL}
\end{figure*}
%%%

Letting $a_m(\xi)=\sqrt{\eta_m(\xi)}\exp i\phi_m(\xi)$ ($m=0,\pm1$), by exploiting  the gauge invariance of Eq.~\eqref{eq:FDHONLSadim}, we reduce the dimensionality of Eq.~\eqref{eq:CMT} to a system of real ODEs in 4 variables: the norm (or wave action) $N_3\equiv\eta_0+\eta_1 + \eta_{-1}$, its fraction in sidebands $\eta\equiv(\eta_1+\eta_{-1})/N_3$, the relative phase $\psi \equiv (\phi_1+\phi_{-1}-2\phi_0)/2$ and the normalized sideband imbalance $\alpha\equiv(\eta_1-\eta_{-1})/N_3$. The reduced system reads as
\begin{subequations}
\begin{align}
\dot{\eta} &= \underbrace{2 \delta_1\Omega\alpha(1-\eta)}_\mathrm{DF} 
			-2 \sigma
		(1-\eta)\left[\eta^2-\alpha^2\right]^\frac{1}{2}\sin 2\psi \\
			\dot\psi &= \frac{\Omega^2}{2} - N_3\left(1 - \frac{3}{2} \eta\right)  
			+ \Sigma \left( s(1-\eta)+ \frac{3}{2}\alpha \right)\\
					\nonumber
			&+ \left\{\sigma\left[\eta^2-\alpha^2\right]^\frac{1}{2}
						- \sigma\eta(1-\eta)\left[\eta^2-\alpha^2\right]^{-\frac{1}{2}}
			 \right.
			 \\
					\nonumber
			&\left.+\Sigma (1-\eta)\alpha\left[\eta^2-\alpha^2\right]^{-\frac{1}{2}} \right\} \cos 2\psi	\\
			\dot{\alpha} &= \underbrace{2\delta_1\Omega(\eta-\alpha^2)}_\mathrm{DF} -
			2\Sigma (1-\eta)\left[\eta^2-\alpha^2\right]^\frac{1}{2}\sin 2\psi
\label{eq:CMT_alpha}			
			\\	
			\dot{N_3} &=\underbrace{2N_3\left(\delta_0 + \delta_1\Omega\alpha\right)}_\mathrm{DF},\label{eq:CMT_E}
\end{align}
\label{eq:CMTreal}
\end{subequations}
with $\sigma=(1- 2 s \Eps\Omega)N_3$ and $\Sigma\equiv 2 \Eps\Omega N_3$. The DF terms $\delta_0$ and $\delta_1$ do not enter in the evolution of phase, but break the constancy of $N_3$ and modify the  coupling between $\alpha$ and $\eta$. As proved in Ref.~\cite{Armaroli2017}, the terms preceded by  $\Sigma$ come from the non-conservation of linear momentum $P \equiv\frac {i}{2}\int_{-\infty}^{\infty}{(a^*_\tau a-a_\tau a^*)\ud \tau}$ in the Dysthe equation for the envelope of the surface elevation. 

Consider a periodic function $a$ of $\tau$. The Fourier transform of $a$, $\hat{a}$ is thus evaluated only at discrete points $m\Omega$. 
Notice that $N_3$ corresponds to the three-wave approximation of the norm $N=\sum_{m=-\infty}^\infty|\hat a(m\Omega)|^2$, while $\Omega\alpha N_3$ corresponds to $P = \sum_{m=-\infty}^\infty m\Omega |\hat a(m\Omega)|^2$.

$\Omega\alpha$ can be regarded as the truncation of $P/N$, i.e.~the spectral mean of $a$, which, see Eq.~\eqref{eq:CMT_alpha}, does not depend directly on $\delta_0$ and shifts according to the sign of $\delta_1$. We recall also that,  frequencies outside the conventional BFI range, where the central mode $\eta_0$ does not exchange energy with its sidebands\cite{Trillo1991c, Armaroli2017},  can be unstable \cite{Brunetti2014}: the energy growth leads to $\eta=1$ and $\alpha=\pm 1$ (according to the sign of $\delta_1$) and the instability rate grows linearly in $\delta_1$ and $\Omega$, see \eqref{eq:CMT_E}.

The ODEs for $N_3$ \eqref{eq:CMT_E}, $N_3\Omega\alpha $, and $\Omega\alpha$ (straightforward to derive from Eqs.~\eqref{eq:CMT_alpha} and \eqref{eq:CMT_E} are, respectively, simpler closed-form relations for the evolution of the quantities $N$, $P$, and $P/N$, which are shown in Appendix \ref{app:conservedquant} to involve higher-order momenta, in an infinite recursion; see Eqs.~\eqref{eq:adimnorm}, \eqref{eq:adimmomentum}, \eqref{eq:adimspmean}.

We can also write
\begin{equation}
\ddot{N_3}-4\delta_0\dot{N_3}+4\delta_0^2N_3= (2\Omega\delta_1)^2\eta N_3 + \Eps f(\eta,\psi,\alpha,N_3)
\label{eq:Eevol}
\end{equation} 
where the function $f(\eta,\psi,\alpha,N_3)$ is rapidly oscillating in $\psi$. Thus the sideband growth is the source of  an overall energy increase which depends only on $\delta_1^2$, provided that the initial buildup of BFI is not suppressed by $\delta_0<0$.

As described in Ref.~\cite{Armaroli2017} and Appendix \ref{app:Hamiltonian}, in the case of $\delta_0=\delta_1=0$, despite the broken integrability of the reduced system ($\alpha$ is not conserved), the perturbed evolution is still quasi-periodic and can be conveniently mapped on the Hamiltonian level sets of the corresponding integrable system.
 {
The heteroclinic structure is summarized in Fig.~\ref{fig:phaseplaneGL}(a); we have three families of solutions (we take hereafter $N_3(0)=1$, $\alpha(0)=0$, $\eta(0)=0.01$, $\Eps=0.1/\sqrt{2}$): (i) the separatrix which connects two saddle points at $\psi\approx\pm \pi/4 \pmod{\pi}$ and corresponds to an infinite period of recurrence (it generalizes the AB of the NLS) (ii) for $\psi(0)=0$,  the orbit is closed and lies inside the separatrix and surrounds the center  $\eta=\eta_c,\,\psi\equiv0\pmod{ 2\pi}$, i.e.~it is period-one; (iii) for  $\psi(0)=\pi/2$, the orbit is open and lies outside the separatrix, i.e.~it is period-two ($\psi$ is shifted by $\pi$ at each recurrence of $\eta$). }
The insets show the full spatiotemporal evolution in the three cases: the drift toward the left and the smaller growth rate (farther focusing point for the AB) are the signatures of HONLS corrections \cite{Zakharov2010,Shemer2013,Zhang2014}.  {The phase-space trajectories of HONLS are topologically equivalent to the trajectories of the low-dimensional system:  a solution of a given family of the former can be continuosly deformed into a solution of the corresponding family of the latter. We recall that the quantitative agreement improves for frequencies closer to the BFI cut-off \cite{Trillo1991c,Armaroli2017}.}
%%%
% Fig. 1 classify solutions
%\begin{figure}[hbtp]
%\centering
%\includegraphics[width=.40\textwidth]{phasespace2}
%\caption{Classification of the HONLS solutions without forcing or damping in the $(\psi/\pi,\eta/N_3)$ plane. Main panel:  level sets of the Hamiltonian associated to the integrable version of \eqref{eq:CMTreal}, see Appendix ?, for $\Eps=0.1$, $N_3=1$, $\alpha=0$. In the insets the solutions of the HONLS corresponding to three different trajectories in the phase plane: (i) period-one, (ii) period-two, (iii) AB. The latter correspond to the separatrix in the phase plane (dashed-dotted blue line level set to be compared to actual orbit in solid blue) and is obtained by using as initial condition the AB solution of NLS with $a=0.32$ (corresponding to the peak BFI gain of the HONLS) at $\xi=-10$. Notice that the peak is reached in the HONLS later than in the NLS, as is well known from the literature.}
%\label{fig:phaseplane}
%\end{figure}
%%%%%%%%%%%%%%%%%%%%%%%%%%%%%

\section{Results and discussion}

\begin{figure}[hbtp]
\centering
\includegraphics[width=.45\textwidth]
{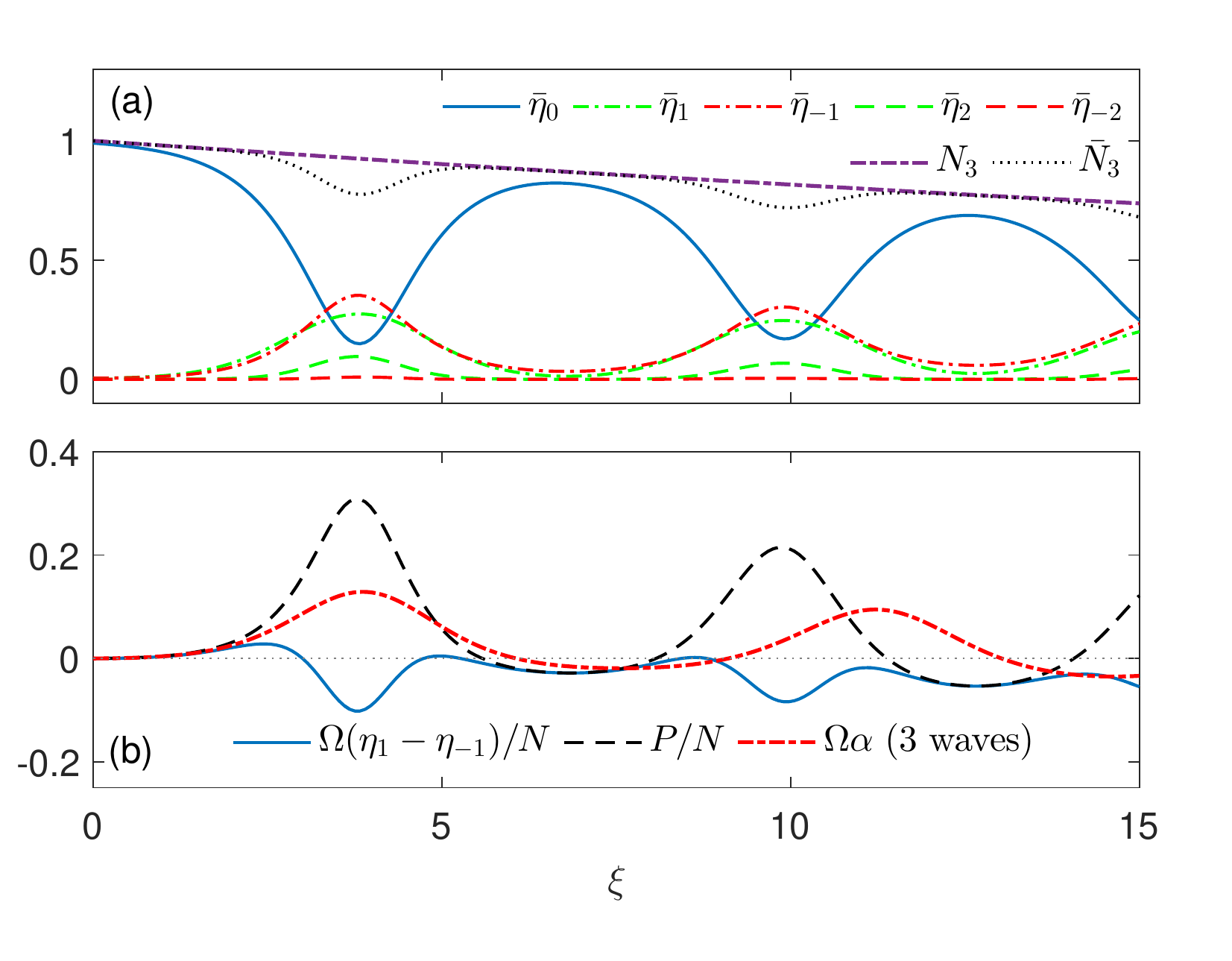}
\caption{Simulated evolution of Eq.~\eqref{eq:FDHONLSadim} and comparison to three-wave truncation Eq.~\eqref{eq:CMTreal}, for $\delta_0=-0.01$, $\delta_1=-0.005$, $\psi(0)=0$. We use the shorthand $\bar\eta_m=|\hat a(m\Omega)|^2$ for the different spectral modes. (a) Evolution of the squared amplitude of the Stokes wave [blue (dark gray) solid line], its unstable sideband pair  ($\pm\Omega$, dash-dotted lines) and second order sidebands ($\pm2\Omega$, dashed lines). Red (dark gray) [green (light gray)] lines represent the low (high) frequency sideband. The black dotted line represent the sum of square amplitudes of Stokes wave and its unstable sidebands ($\bar N_3=\sum_{m=-1}^1{\bar\eta_m}$) to be contrasted to $N_3$ obtained from Eq.~\eqref{eq:CMTreal} [purple (dark-gray) thick dash-dotted line]. (b)  Corresponding  relative imbalance of sideband amplitudes $\Omega(\bar\eta_1-\bar\eta_{-1})/N$ (solid blue) and spectral shift $P/N$ (dashed black),  the truncated version of which,  $\Omega\alpha$, is shown by the dash-dotted red line. 
%The green dotted line serves as eye guidance to identify the zero of peak spectral shift. 
}
\label{fig:negneg}
\end{figure}

\begin{figure}[hbtp]
\centering
\includegraphics[width=.45\textwidth]{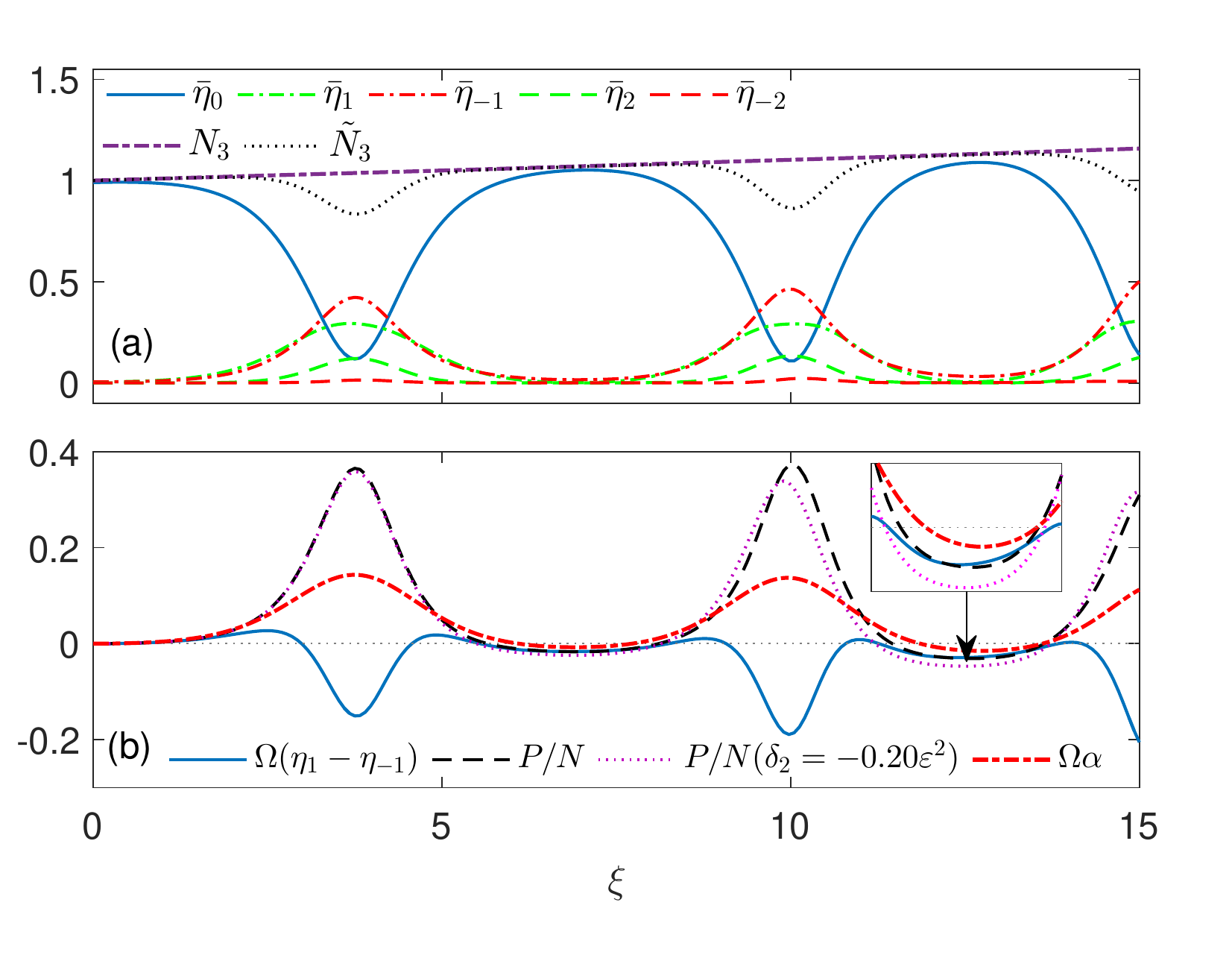}
\caption{Same as Fig.~\ref{fig:negneg} with $\delta_0=0.005$,  $\delta_1=-0.0025$, and $\psi(0)=\pi/2$. In panel (b) we add the simulation result for $P/N$ (dotted purple line) obtained by adding a parabolic damping $\delta_2 = -0.20\epsilon^2$ The inset finally represents a detail of the zero crossing of $P/N$ and $\Omega\alpha$.}
\label{fig:posneg}
\end{figure}

\begin{figure}[hbtp]
\centering
\includegraphics[width=.45\textwidth]{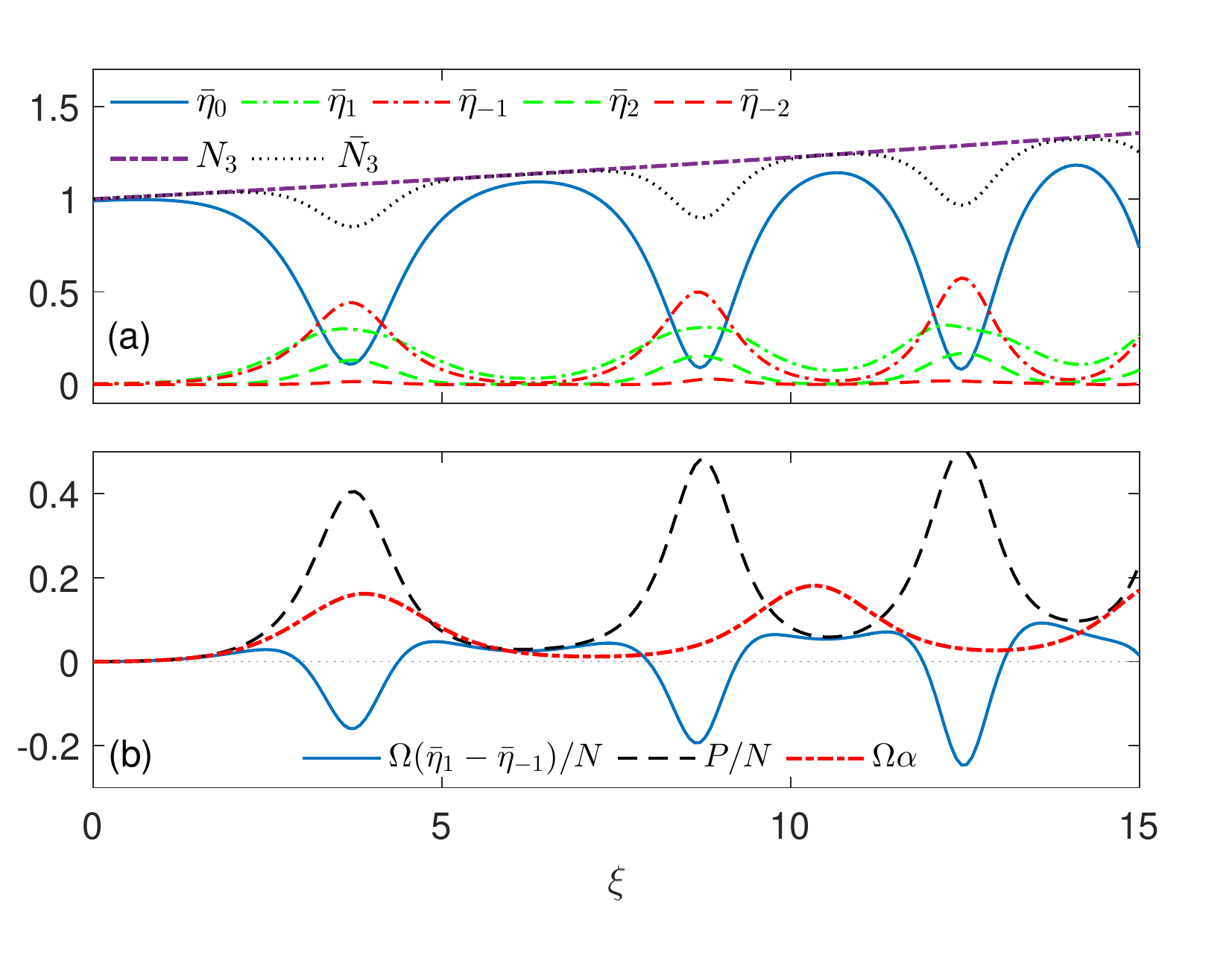}
\caption{Same as Fig.~\ref{fig:posneg} with $\delta_0=0.01$ and $\delta_1=0.0025$.}
\label{fig:novisc}
\end{figure}

In this section, we show that the three-wave truncation is a good approximation for the DF-HONLS, Eq.~\eqref{eq:FDHONLSadim}, and that the three damping/forcing regimes mentioned above can be classified according to the shift of spectral mean and the topology of the solution in the phase plane. 
%We study the solutions for $N_3(0)=1$, $\alpha(0)=1$, $\eta(0)=0.01$ and $\Eps=0.1$. 
We checked numerically (not shown) that the value of $\Eps$ chosen above prevents wave-breaking, even under the action of wind. We use as a criterion that the amplitude of the peaks at the focusing point of the recurrence cycles never exceeds $|a|=3$ \cite{Tulin1999,Toffoli2010a}.  In Fig.~\ref{fig:phaseplaneGL}(b-d), the reader will find the  representation of trajectories in the   phase plane ($\psi/\pi,\eta$): the level sets are computed at fixed $N_3$. In Figs.~\ref{fig:negneg}--\ref{fig:novisc} we present the simulated dynamics of the energy exchange among the carrier wave, the pair of unstable sidebands and their four-wave mixing product at $\pm 2 \Omega$ [panels (a)]; we define $\bar\eta_m\equiv |\hat a(m\Omega)|^2$; we focus here on the three-wave approximation $N_3$ to simulated $\bar N_3 = \bar \eta_0 + \bar \eta_1 + \bar \eta_{-1}$ in order to assess the accuracy of our truncation. 

We omit to include the simulated $N$, because it coincides in each case to $N_3$ from the truncated model. Panels (b) show the spectral shift according to its possible definitions---unbalance (related to spectral peak) and spectral mean: in particular $\Omega\alpha$ is compared to $P/N$.

%%%
% Fig. 3 damping and downshift

%%%
% Fig. 4 forcing and upshift

%%%
% Fig. 5 forcing and upshift

We start by generalizing the findings of Ref.~\cite{Kimmoun2016}, where the evolution of an AB is studied under the action of small viscosity; we add here a weak wind to partly compensate the viscous damping: we take  {$\delta_0=-0.01$ and $\delta_1=-0.005$}. The initial conditions are the same as mentioned at the end of the previous section, with $\psi(0)=0$.
In Fig.~\ref{fig:negneg}(a) we observe several cycles of quasi-recurrence. As expected  $\bar\eta_{-1}$ grows more than $\bar\eta_{1}$. Instead of reconverting back all the energy to $\bar\eta_0$, the former does not return to zero and gradually increases from one cycle to the other, the latter oscillates as in the undamped case. The total energy decays, so at each recurrence cycle the conversion is weaker and the period is longer. $\bar N_3$ deviates only slightly (about 15$\%$ at maximum conversion) from  $N_3$: this confirms the soundness of the truncation. Specifically, $\bar\eta_2$ acquires less than 10$\%$ of the total energy and its opposite at $\bar\eta_{-2}$ is one order of magnitude smaller, consistently with the temporary mean upshift caused by HO corrections. 
%Notice that $N_3$ matches almost perfectly with the simulated evolution of $N$. 
The downshifting trend is apparent in Fig.~\ref{fig:negneg}(b): the peak downshift and mean upshift due to HO corrections are temporary, but are not exactly reverted at each cycle; the average of $P/N$ points downward and the three-wave $\Omega\alpha$ approximation represents qualitatively well this trend both on the short and long $\xi$ scales:  {at each breathing period (which the truncation overestimates by about 25$\%$) a temporary upshift is observed. Despite a lower maximum value  (by half) for $\Omega\alpha$ than for $P/N$, they agree remarkably well in the decrease of the value attained at the end of each cycle with respect to its beginning.} 
In Fig.~\ref{fig:phaseplaneGL}(b), we observe that the trajectory in the phase plane  crosses the separatrix during the first recurrence cycle and is attracted to a period-two orbit, characterized by a longer recurrence period. This is a consequence of energy damping.

% Fig.~\ref{fig:negneg}(a). As the low-frequency modes experience stronger losses, the overall effect is that the temporary spectral peak downshift and the mean upshift oscillates less from a focusing period to the next, panel (b). The trend points toward an effective downshift. Our three-wave approximation  captures well this phenomenon: both $N$ in panel (a) and $P/N$ in panel (b) are correctly approximated by the three-wave truncation (the sum $\eta_0+\eta_{1}+\eta_{-1}$ matches $E$ computed from the three-wave system and $\alpha/E$ is qualitatively representing the trend of $P/N$ both on the short and long $\xi$ scales). Notice in panel (a) that less than 10$\%$ of the energy is converted to $2\Omega$ and $-2\Omega$ is always one order of magnitude smaller, consistently with the temporary growth of $P/N$. The rate of energy conversion  is also well represented on the phase-plane Fig.~\ref{fig:phaseplaneGL}(b). The overall energy damping implies that the approximation behaves better and better during the evolution. The only remarkable discrepancy between the full model and the three-wave truncation is in the period of recurrence, analogously to what is observed in the NLS and HONLS cases \citep{Trillo1991c,Armaroli2017}. 

Thus our representation provides an insight to how a small viscosity causes a generic solution inside or on the separatrix to cross into an open orbit with longer recurrence periods. From the definition of $\delta_0$ and $\delta_1$, we can affirm that a small viscosity is inextricably associated to a downshift of the spectral mean.

The second example is  {$\delta_0=0.005$, $\delta_1=-0.0025$}, i.e.~when the wind dominates the constant part of the forcing but is overruled by viscosity in the linear dispersive contribution.  {These values are sufficiently small to guarantee the validity of the model \eqref{eq:DFdisp} and the experimental  feasibility, see below and App.~\ref{app:DFdisp}}. We start from a period-two orbit, $\psi(0)=\pi/2$.  The energy grows and is concentrated in $\bar\eta_{-1}$, which does not return to zero at each recurrence cycle, while $\bar\eta_0$ is less affected, see Fig.~\ref{fig:posneg}(a). $\bar\eta_1$ oscillates instead with a constant amplitude and is back to zero at each breathing period. Notice that  the period of recurrence is shortened at each cycle. $\bar\eta_{\pm 2}$ oscillate quasi periodically as in the HONLS case. Thus our truncation is still valid here, because only 3 sidebands take part in the evolution and the others are forced to follow them, see the small deviation of $\bar N_3$ from $N_3$ (or equivalently $N$). Fig.~\ref{fig:phaseplaneGL}(c), we observe that the period-two solution is thus attracted to oscillate around a point in the phase-plane which corresponds to the original center of the HONLS $(0,\eta_c)$ (see Appendix \ref{app:Hamiltonian}).  {
We choose the phase of the initial conditions outside the separatrix, such that a crossing to the other orbit is observed, demonstrating the attraction. An initial condition inside the separatrix would stay inside and the attraction would not be apparent.}
  {In panel (b) we see that, analogously to  the previous case, the average spectral shift points downwards but it is weaker because of smaller value of $|\delta_1|$. The truncated model still predicts,  in a qualitative way, both the monotonic increase of $N_3$ and the oscillating decrease of $P/N$. Both $\delta_0$ and $\delta_1$ contribute to a net forcing; the additional energy is concentrated on $\eta_{-1}$. This example is the only one where a parabolic correction of Eq.~\eqref{eq:DFdisp} has an effect: the period of recurrence is $2\%$ shorter and the accumulated downshift about 50$\%$ larger (around $\xi=12.5$), but not qualitatively different from $\Omega\alpha$, see the dotted line in Fig.~\ref{fig:posneg}(b).}

It can be verified that these first two examples correspond to the results presented in Ref.~\cite{Touboul2010}. 

Finally we explore the case where viscosity is negligible with respect to wind forcing  at both orders, i.e.~$\delta_0=0.01$ and $\delta_1=0.0025$. The orbit is attracted from a period-two [$\psi(0)=\pi/2$] to a period-one orbit  {(as in the previous case, every solution ultimately crosses the separatrix)}, see Fig.~\ref{fig:phaseplaneGL}(d); the period of recurrence is again shortened at each cycle, Fig.~\ref{fig:novisc}. The energy is transferred to larger $\bar\eta_{-1}$ oscillations and an average growth of $\bar\eta_1$, that  does not return to zero, Fig.~\ref{fig:novisc}(a). $\bar\eta_{\pm2}$ still evolve almost unaffected. While the oscillations of the spectral peak are wider, the spectral mean shows a clear upshifting trend Fig.~\ref{fig:novisc}(b).  This is consistent with the findings of Ref.~\cite{Eeltink2017} and shows the importance of properly defining the spectral shift: the apparent downshift of the peak is only temporary and  corresponds to the focusing points as predicted by the conservative HONLS, but a permanent upshift of the mean is caused by wind. 

We conclude this section with three remarks. First, even for Fig.~\ref{fig:novisc}, as far as the accuracy of the three-wave truncation is concerned, $N_3$ superimposes perfectly to $N$ and, from the plot of $\bar N_3$, at most 25$\%$ of the energy is converted to higher-order sidebands. We checked (not shown) that a five-wave version of Eq.~\eqref{eq:CMT} better matches with the solution of \eqref{eq:FDHONLSadim} (from  50$\%$ to  30$\%$ underestimation of the  peak  $P/N$ and from 25$\%$ to 10$\%$ overestimation of the period of recurrence, analogously to what is observed in the NLS and HONLS cases \cite{Trillo1991c,Armaroli2017}), but at the expense of the availability of a simple phase-plane analysis.

Second, the need of a very long wave-tank, in order to observe many recurrence cycles, is the main limitation to  experimentally discriminate the three regimes; the longest tank we are aware of is 200 m long\cite{Kimmoun2016}. 
As an example, for $0<\xi<15$,   $L_0=13.33$ m, and the resulting $k_0 =7.50\, \mathrm{m^{-1}}$. The resulting time scaling factor is then $T_0=1.649$ s.  {We can derive that for the example of Fig.~\ref{fig:negneg}, $\Gamma_m=1.78\times 10^{-3} \; \mathrm{[s^{-1}]}$ and $\nu=1.17\times 10^{-5} \; \mathrm{[m^{2}/s]}$, for Fig.~\ref{fig:posneg}, $\Gamma_m=5.18\times 10^{-3} \; \mathrm{[s^{-1}]}$ and $\nu=2.11\times 10^{-5} \; \mathrm{[m^{2}/s]}$ and finally $\Gamma_m=1.26\times 10^{-3} \; \mathrm{[s^{-1}]}$ and $\nu=1.77\times 10^{-6} \; \mathrm{[m^{2}/s]}$ for Fig.~\ref{fig:novisc}. The corresponding wind speeds are estimated to 2.78, 4.38, 2.40 [m/s]} respectively \cite{Miles1957}.

Third, we should not forget that in  Eq.~\eqref{eq:FDHONLSdim} the  viscous contribution to boundary conditions is derived in the linear limit for small $k_0$ (gravity waves) and small viscosity \cite{Wu2006,Dias2008} and the model of wind may be inaccurate if we enter in a turbulent regime. 

\section{Conclusions}

We applied a three-wave truncation to study the recently derived forced/damped high-order nonlinear Schr{\"o}dinger equation for one dimensional gravity waves in deep water under the concurrent effect of wind and viscosity.  {Damping and forcing are modeled in the  frequency domain by a homogeneous plus a linear term.}
The low-dimensional approach proves effective to approximate the evolution of  one unstable mode evolution. $N_3$ perfectly matches to the norm $N$ and $\Omega \alpha$ represents a good approximation of the spectral mean $P/N$, with the advantage of evolving according to simple real closed-form ODEs.

% We derived a closed form for the evolution of norm $N_3$ and spectral mean $\Omega\alpha$.

These results allow us to classify the solutions of the damped/forced HONLS according to two complementary parameters: the shift of the spectral mean and  the topology of orbits in the phase plane of a low dimensional dynamical system.  
 {A \emph{downshift} is compatible with a period-one attractor, if wind dominates the homogeneous contribution (in frequency), or period-two attractor, if viscosity dominates at both orders. The elusive spectral \emph{upshift} coexists, instead, only with the attraction to a period-one orbit, if wind forcing dominates both constant and linear forcing terms.}
%Frequency \emph{upshift} necessarily coexists with a growth of energy and a period-one attractor, while \emph{downshift} can be observed both with  period-one and period-two attractors, depending on the relative magnitude of wind-forcing and viscosity. 

The proposed analysis and classification  {is not model-specific and will allow to better understand the role of wind-forcing and viscosity on water waves in very long tanks and to guide the improvement of the available propagation models.}

\begin{acknowledgments}
We acknowledge the financial support from the Swiss National
Science Foundation (Project No.~200021-155970). We would like to thank John D.~Carter for fruitful discussions. 
\end{acknowledgments}
%\bibliographystyle{aip}
%\bibliography{hydro,rogue}

\appendix
 {
\section{A simple linear derivation of damping/forcing terms}
\label{app:DFdisp}
We present here a simple derivation of the damping and forcing terms in Eq.~\eqref{eq:FDHONLSdim}, which does not require to apply the method of multiple scales, as in Ref.~\cite{Eeltink2017}, but a simple linear argument. It is well known and easy to verify that the linear part of the evolution equation is derived from the dispersion relation $\omega(k)$, while the propagation equation of Eq.~\eqref{eq:FDHONLSdim} is based on $k(\omega)$; this explains the simple form of HONLS not involving dispersion corrections of order higher than the second.
\subsection*{Viscous damping} 
We start from the well-known implicit equation for the dispersion relation under the action of viscuous damping\cite{lamb1945hydrodynamics,Dias2008}
\[
\left(2-i\frac{\omega}{\nu k^2}\right)^2 + \frac{g}{\nu^2 |k|^3} = 4\left(1-i\frac{\omega}{\nu k^2}\right)^\frac{1}{2}.
\]
The conventional dispersion relation, i.e.~$\omega(k)\approx\sqrt{g k } - 2 i \nu k^2$, is obtained by neglecting the right-hand side of the implicit equation.
In the limit of small viscosity, $\nu k^2/\omega \sim \mathcal{O}(\epsilon^2)$, we can easily invert the dispersion relation at order $\mathcal{O}(\epsilon)$ as $k(\omega)\approx\frac{\omega^2}{g} + 4i \nu \frac{\omega^5}{g^3}$. The damping can be Taylor-expanded around $\omega_0$ as $\frac{4 \nu}{g^3} (\omega_0^5 + 5\omega_0^4 \Omega + 10\omega_0^3 \Omega^2+\ldots$), where we identify the first two terms with the homogeneous and linear damping in Eq.~\eqref{eq:FDHONLSdim}. The parabolic damping term is thus $\tilde\delta_2^D =  -\frac{40 \nu \omega_0^3}{g^3}$ in the dimensional equation and $\delta_2^D = -\frac{20\nu k_0^2}{\omega_0}$.
\subsection*{Wind forcing}
According to Miles mechanism\cite{Miles1957}, the extra pressure due to wind forcing divided by the fluid density can be written as $\frac{P}{\rho}=\gamma \eta_x$. It is possible to identify $\gamma = \frac{\omega_0}{k_0^2}\Gamma_m$, where $\Gamma_m$ is the growth rate at frequency $\omega_0$ defined above. The dispersion relation is now the solution of the algebraic equation $-\omega^2 + g |k| + i \gamma k^2 = 0$, where we neglect viscosity. For $\gamma k^2\sim \mathcal{O}(\epsilon^2)$, we write $k(\omega)\approx\frac{\omega^2}{g} - i \gamma \frac{\omega^4}{g^3}$. The forcing is thus expanded as $\frac{\gamma}{g^3}(\omega_0^4 + 4 \omega_0^3\Omega + 6 \omega_0^2 \Omega^2+\ldots)$. As above, the first two terms correspond to homogeneous and linear (derivative) forcing in  Eq.~\eqref{eq:FDHONLSdim}. The parabolic correction to forcing, $\tilde\delta_2^F = \frac{6\gamma  \omega_0^2}{g^3}$, is adimensionalized as $\delta_2^F = \frac{3\Gamma_m}{\omega_0}$.
\subsection*{Net parabolic damping/forcing}
The balance of the two parabolic terms of damping and forcing allows us to extend Eq.~\eqref{eq:DFdisp} to
$$
g(\Omega) = \delta_0 + \delta_1 \Omega + \delta_2 \Omega^2,
$$ 
with $\delta_2 = \delta_2^F+\delta_2^D = T_0 \epsilon (3\Gamma_m - 20 \nu k_0^2)$. The choice of $\delta_{0,1}$ in the text  limits the total damping of the unstable sidebands to 5$\%$, i.e. $\exp(-\delta_2 \Omega_M\xi) > 0.95$ at the end of the wave-tank. 
}

\section{Evolution of moments}
\label{app:conservedquant}
We mentioned in the text that the HONLS conserves only the norm $N$ in the surface-elevation formulation, while it conserves also $P$ in the velocity potential formulation. We recall also that in this latter case it conserves also the Hamiltonian $E \equiv E_0-\epsilon E_1$, with $E_0=\frac{1}{2}\int_{-\infty}^{\infty}{\left(|a_\tau|^2- |a|^4\right) \ud\tau}$ and $E_1=i\int_{-\infty}^{\infty}{|a|^2\left[2(a_\tau^* a - a_\tau a^*) - \mathcal{H}[|a|^2]_\tau\right] \ud\tau}$. Properly canonical versions were derived, see \cite{Gramstad2011}.

We report here briefly the evolution of the most important quantities (norm, momentum, spectral mean) in the damped-forced model, in order to better compare to the approximated results reported in the text, Eqs.~\eqref{eq:CMT_alpha} and \eqref{eq:Eevol}.

From Eq.~\eqref{eq:FDHONLSdim} it is easy to obtain, for the dimensional quantities,
\begin{equation*}
	\frac{\ud \tilde N}{\ud \tilde x} = \frac{2k_0}{\omega_0}\left( \Gamma_m-4\nu k_0^2\right) \tilde N
+ \frac{8k_0}{\omega_0^2}\left( \Gamma_m   -5\nu k_0^2\right) \tilde P
\end{equation*}
where $\tilde N\equiv\int_{-\infty}^{\infty}{|A|^2\ud \tilde t}=A_0^2T_0 N$ and  $\tilde P \equiv-\mathrm{Im}\left\{\int_{-\infty}^{\infty}{A^*_{\tilde t} A \,\ud \tilde t}\right\}=A_0^2 P$
\begin{equation*}
	\frac{\ud \tilde P}{\ud  \tilde x} = \frac{2k_0}{\omega_0}\left( \Gamma_m-4\nu k_0^2\right) \tilde P + 
\frac{8k_0}{\omega_0^2}\left( \Gamma_m   -5\nu k_0^2\right) \tilde Q - \frac{4 k_0^3}{\omega_0} \tilde R
\end{equation*}
where $\tilde Q \equiv \int_{-\infty}^{\infty} |A_{\tilde t}|^2 \,\ud \tilde t$ and $\tilde R \equiv \mathrm{Im} \left[\int_{-\infty}^{\infty} |A|^2 A^* A_{\tilde t \tilde t} \, \ud \tilde t \right]$.
The spectral mean $\tilde \omega_m\equiv \frac{\tilde  P}{\tilde N}$ thus evolves according to
\begin{equation}
	\frac{\ud}{\ud \tilde x} \tilde \omega_m \equiv \frac{8k_0}{\omega_0^2}\left( \Gamma_m   -5\nu k_0^2\right) \left(\frac{\tilde Q}{\tilde N}-\tilde \omega_m^2\right) - \frac{4 k_0^3}{\omega_0} \frac{\tilde R}{ \tilde N}  
	\label{eq:spmean}
\end{equation}
The first term in the right-hand side of Eq.~\eqref{eq:spmean} is the effect of the fourth-order contribution of wind and viscosity, the second one appears because of the symmetry breaking in the surface-elevation HONLS.

We can write the same equations from Eq.~\eqref{eq:FDHONLSadim} as well. They read as
\begin{equation}
\dot{N} = 2\delta_0 N + 2\delta_1 P
\label{eq:adimnorm}
\end{equation}
which corresponds to Eq.~\eqref{eq:CMT_E},
\begin{equation}
\dot {P} = 2\delta_0 P + 2\delta_1 Q - 4 \Eps R
\label{eq:adimmomentum}
\end{equation}
where we use $Q=T_0/A_0^2 \tilde{Q}$ and $R=T_0/A_0^4 \tilde{R}$.
% These quantities, defined for the adimensional system in the same way than above for dimensional one, can be expressed, in the three-wave approximation, explicitly; compare to Eq.~\eqref{eq:CMTreal}. 

The norm evolves according to 
\[
	\ddot{N} -4\delta_0 \dot{N} + 4\delta_0^2 N = 4\delta_1^2 Q - 8 \Eps \delta_1 R,
\]
the  form of which is identical to Eq.~\eqref{eq:Eevol}, where we identify the forcing proportional to $\delta_1^2$ and the small oscillating terms proportional to $R$.

Finally we derive from Eq.~\eqref{eq:spmean} the evolution of the spectral mean ($\omega_m\equiv P/N= \tilde\omega_mT_0$)  in normalized form:
\begin{equation}
\dot{\omega}_m = 2\delta_1 \left(\frac{Q}{N} - \omega_m^2\right)-4\Eps\frac{R}{N},
\label{eq:adimspmean}
\end{equation}
which can easily be put in one-to-one correspondence to Eq.~\eqref{eq:CMT_alpha}. 

We conclude that, while the space evolution of each quantity involves higher-order momenta,  the truncation gives us more transparent and closed form relationships among them.

\section{Hamiltonian function of three-wave truncation}
\label{app:Hamiltonian}
In the absence of forcing/damping, the HONLS for the velocity potential can be written without the term $2a^2\pder{a^*}{\tau}$ in Eq.~\eqref{eq:FDHONLSadim}. We showed in Ref.~\cite{Armaroli2017} that the three wave truncation conserves $N_3$  and $\alpha$, thus the system is reduced---without loss of generality we let $N_3=1$---to a one degree-of-freedom system in the conjugate variables $(\psi,\eta)$ with Hamiltonian function
($\dot{\eta}=\frac{\partial H}{\partial \psi}$, $\dot{\psi}=-\frac{\partial H}{\partial \eta}$), with Hamiltonian 
%{[\color{red} If we keep $U$, we can extend to wind, if we put $N_0=1$, it's not possible]}
\begin{multline}
H(\psi,\eta) = -\left(\frac{\Omega^2}{2}  -\sigma  + \epsilon 4\Omega\alpha\right)\eta - \frac{3}{4}\eta^2 \\+  \epsilon  s\Omega\eta^2 + \sigma(1-\eta)\left[\eta^2-\alpha^2\right]^\frac{1}{2}\cos 2\psi
\label{eq:hamiltonian}	
\end{multline}
with $\sigma=(1-2s\epsilon\Omega)$.
The extrema of $H$ correspond to the fixed point of the dynamics. For $\alpha=0$ it is easy to derive the saddle point from which the separatrix emanates $\left(\psi = \pm \frac{1}{2}\cos^{-1}\left(\frac{\Omega^2}{2\sigma}-1\right),\;\eta=0\right)$  and the center $\left(\psi\equiv 0\pmod{\pi},\;\eta_c= \frac{4\sigma - \Omega^2}{3+4\sigma-4\epsilon |\Omega|}\right)$. 

\end{document}